\documentstyle[11pt,pasp,twoside,epsf]{article}
\def\edcomment#1{\iffalse\marginpar{\raggedright\sl#1\/}\else\relax\fi}
\reversemarginpar

\def\*{$^{*}$}

\begin{document}
\title{Observation of solar flares through the ART-P telescope side shield}
\author{Lutovinov A., Pavlinsky, M., Grebenev S.}  
\affil{Space Research Institute, Profsoyuznaya 84/32, 117810 Moscow, Russia}

\begin{abstract}
Some preliminary results of observations of six solar flares though the
ART-P telescop side shield in 1990-1992 are presented.
\end{abstract}

\section{Introduction}

The X-ray imaging telescope ART-P was one of the two main instruments on
board the Granat Observatory, which was launched in 1989. The telescope
consists of four coaxial, completely independent modules; each module
includes a coded mask, a collimator and a position-sensitive detector
(energy band is $\sim$3-100 keV) with a geometric area of 625 cm$^2$. The
detector equipped with the vessel, window, the analog and digital
electronics units. The vessel is made of 4 mm thickness of the titan, which
covers a 1.5 mm of the tin as a shield. The collimator has a dual role, it
acts as a support structure for the thin window against the internal
pressure of the detector and it limits the field of view. The collimator
cell geometry is chosen to be square with 6$\times$6 mm. The material of the
collimator is a copper with thickness of 200 $\mu$m.

\section{Observations}

The analysis of the archive data obtained with ART-P in 1990-1992 has shown
that more than 100 X-ray bursts and flares were detected during this period
(Grebenev et al., 2001). Most of them were identified as type I X-ray bursts
from low-mass X-ray binaries, but also we found six unusual events.
Comparison of them with the PHEBUS/GRANAT instrument data, obtained in the
80-600 keV energy band (Terekhov et al., 1996), has allowed us to conclude
that the telescope ART-P detected solar flares passed through its side
shield. Time profiles of several of them were practically the same in a wide
energy range (3-600 keV), whereas profiles of other flares in high and low
energy bands differed considerably (Fig.1).

\begin{figure}[t]
\plottwo{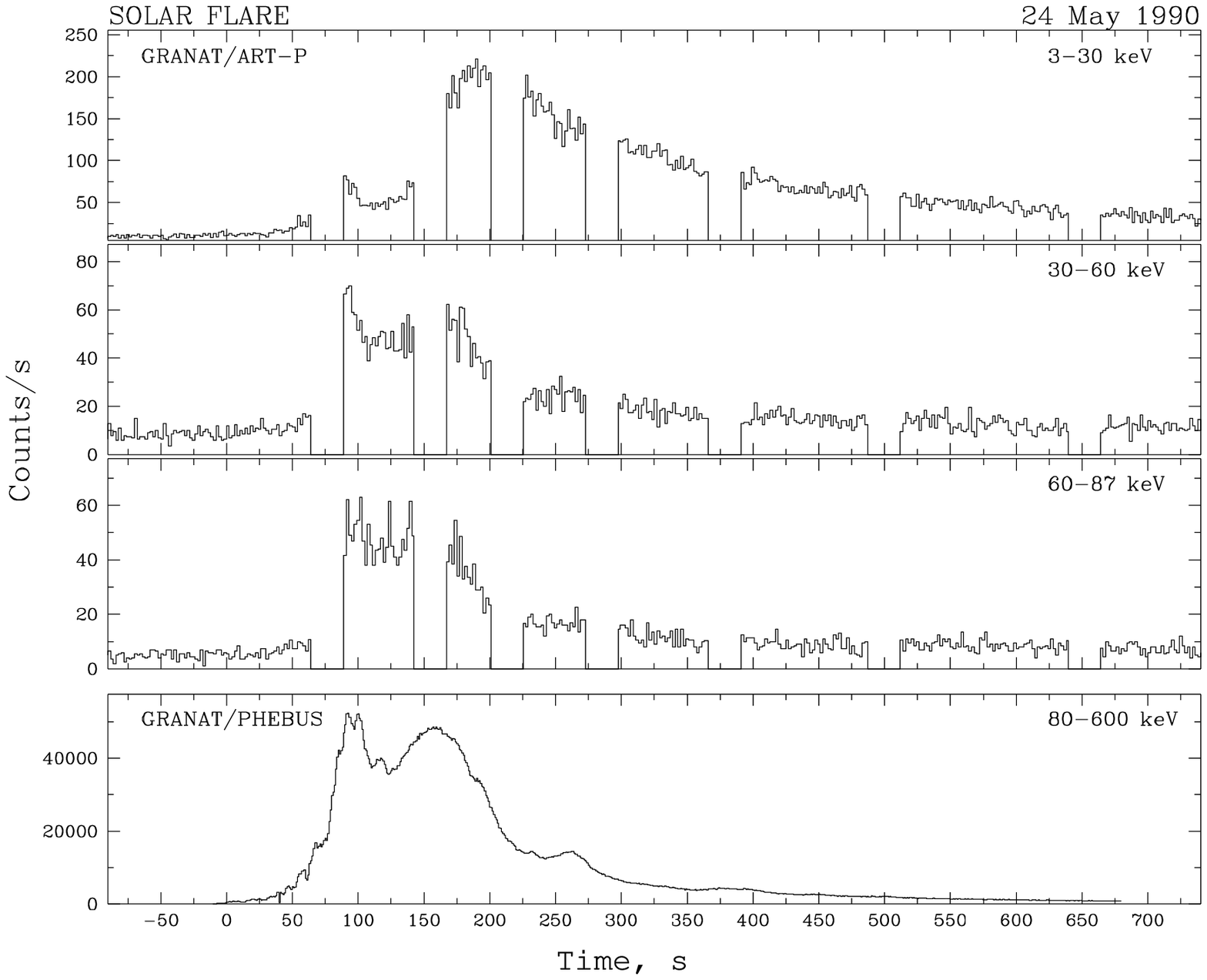}{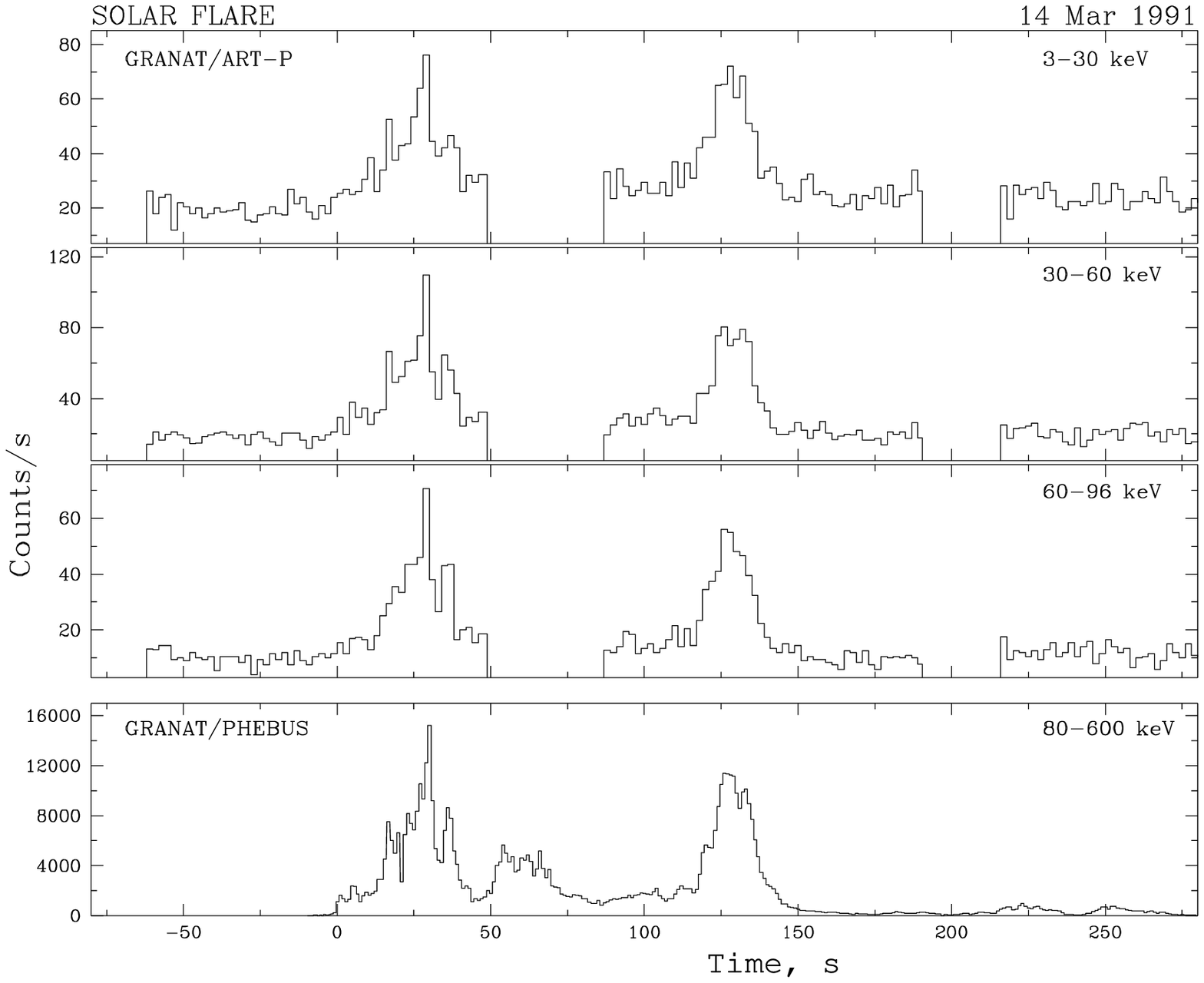}
\vspace{-16mm}\caption{Energy dependence of the flares profiles as derived from
ART-P and PHEBUS data for different types of flares.}
\end{figure}

The solar flares observed by PHEBUS instrument and ART-P telescope on May
24, 1990 was of our special interest. This event is associated with a limb
flare (N33 W78), where GeV protons were observed in interplanetary space and
the largest increase due to solar neutrons was recorded at ground
levels. The spectrum obtained with ART-P during this flare is shown in Fig.2
({\it left panel}). As is obvious from the figure that the powerful emission
feature near 9 keV is present in the flare spectrum. This feature was
interpreted as the K-fluorescence line produced by the copper. Such point of
view is supported by the {\it right panel} of Fig.2, in which the counts
distribution along one of the detector axis are shown. The 6mm-collimator
structure is obvious from this figure.

The detailed analysis of other solar flares detected by the ART-P telescope
and the evaluating of possibility to detect solar flares or powerful
gamma-ray bursts during the INTEGRAL and SRG missions now in progress.

\begin{figure}[h]
\vspace{-60mm}\plotone{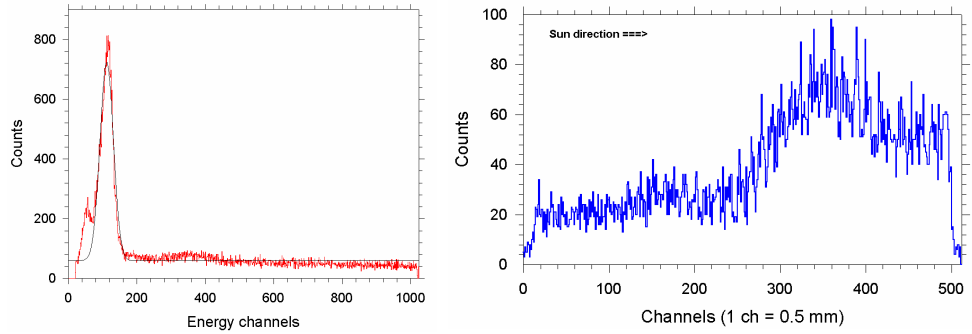}
\vspace{-76mm}\caption{Spectrum of the solar flare observed on May 24, 1990.
Solid line is represent the best-fit approximation by Gaussian with a centroid
energy $E\simeq112$ cnl, what corresponds to $\sim8.7$ keV ({\it left
panel}). Counts distribution along the Sun oriented detector axis ({\it
right panel}).}  
\end{figure}

\end{document}